\documentclass[11pt]{article}
\usepackage{moriond,epsfig}

\def\be{\begin{equation}}
\def\ee{\end{equation}}
\def\bea{\begin{eqnarray}}
\def\eea{\end{eqnarray}}

\begin{document}
\title{SPIN TORQUE IN A MAGNETIC TRILAYER COUPLED TO A SUPERCONDUCTOR}
            
\author{\underline{XAVIER WAINTAL}$^1$,
PIET W. BROUWER$^1$}

\address{$^1$Laboratory of Atomic and Solid State Physics,
Cornell University, Ithaca, NY 14853-2501}

\maketitle\abstracts{A ferromagnet--normal-metal--ferromagnet trilayer
does not conserve spin current and as a result, the conduction
electrons can create a torque on the magnetic moments. 
When the trilayer is connected to a superconducting electrode, 
the spin torque created by a current flowing perpendicularly 
to the trilayer can drive the system to
a configuration where the two magnetic moments are perpendicular to
each other. Here we argue that in contrast to the non-equilibrium
torque, the equilibrium torque (or magnetic exchange interaction)
does not single out the perpendicular configuration. We calculate the
equilibrium torque for a one-dimensional lattice model and find that
it is sensitive to the presence of the superconductivity.}

\section{Introduction}
Spin current is a useful concept to describe the effect of the
conduction electrons on the dynamics of the magnetic moments in
a ferromagnet--normal-metal--ferromagnet
trilayer.~\cite{slon1,slon2,bauer,wmbr} The interaction with the local
exchange field causes the potential felt by the conduction electrons 
to be spin dependent. In the presence of a spin-dependent potential,
the spin current carried by the conduction electrons is
not conserved in the ferromagnetic layers. Since the total spin of the system
is conserved, the missing spin current must be accounted for as a 
torque acting on the magnetic moments of the ferromagnets.\cite{slon1} 
The torque from non-conservation of the equilibrium spin currents is
known as the magnetic exchange (or RKKY) interaction.\cite{slon2}
(There is a direct analogy between the equilibrium spin current flowing
in presence of an non zero angle between the two magnetic moments and 
the persistent current flowing through a mesoscopic ring in presence 
of an Aharonov-Bohm flux.) 
An additional torque is exerted out of equilibrium when a
current is passed through the system. Experimentally, 
this non-equilibrium torque has been shown to be large enough to switch 
the relative orientation
of the two magnetic moments of the two magnetic layers from
parallel to antiparallel to each other.~\cite{exp}
In this paper, we report on our investigation of these torques 
when the trilayer is connected to 
a superconducting electrode. The non-equilibrium torque has been
considered in Ref.\ \ref{wb} where it is shown that the torque can
drive the system to a configuration where the two magnetic moments 
are perpendicular to each other. Here, we address the question
whether a similar effect can be expected for the (equilibrium)
magnetic exchange interaction.

\section{Definition of the spin torque}

\begin{figure}
\begin{center}
\psfig{figure=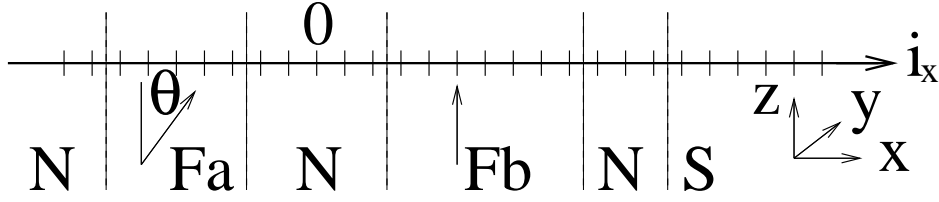,height=1.0in}
\end{center}
\caption{Schematic of the
Ferromagnet--Normal-metal--Ferromagnet
(FNF) trilayer under consideration. Two ferromagnetic layers,
F$_a$ and F$_b$, separated by a normal metal spacer
(N), are connected to one normal-metal
contact (left), and one superconducting contact (S, right). 
(A normal-metal spacer is added between F$_b$ and S.)
The arrows indicate the directions of the local 
exchange fields, $\vec h_a$ ($\theta_i=\theta$ for $i\in F_a$) and 
$\vec h_b$ ($\theta_i=0$ for $i\in F_b$).
\label{fig:schema}}
\end{figure}

Following Ref.\ \ref{wb}, we consider a trilayer system 
consisting of two ferromagnetic layers F$_a$ and F$_b$ with a
normal metal (N) spacer and one superconducting contact (S),
see Fig.\ \ref{fig:schema}. The magnetic moment of the layer F$_b$ 
adjacent to S is considered fixed, e.g., by anisotropy forces.
In a lattice formulation, the trilayer is described by the
the Bogoliubov-De Gennes (BdG) equation~\cite{degennes}
\begin{equation}
H_i \Psi_i + t \sum_{\langle i j\rangle}\Psi_{j}=\epsilon \Psi_{i}.
\label{eq:BdG}
\end{equation}
Here $\Psi_i$ is the quasiparticle wavefunction amplitude on site 
$i=(i_x,i_y,i_z)$, which has an electron-hole as well as a spin index, 
$\Psi_i=(\psi_i^{e\uparrow},
 \psi_i^{e\downarrow},\psi_i^{h\uparrow},\psi_i^{h\downarrow})^T$. 
Further, $t$ is the hopping amplitude, $\sum_{\langle i j\rangle}$ stands
for the sum over nearest neighbors and,
\begin{equation}
H_i=\left(
\begin{array}{cccc}
v_i-\epsilon_F + h_i \cos\theta_i & h_i \sin\theta_i       & 0 & \Delta_i \\
h_i \sin\theta_i       & v_i-\epsilon_F - h_i \cos\theta_i  & -\Delta_i & 0 \\
      0       & -\Delta_i^* & -v_i+\epsilon_F - h_i \cos\theta_i &-h_i\sin\theta_i\\
   \Delta_i^* & 0 & -h_i \sin\theta_i       & -v_i+\epsilon_F + h_i \cos\theta_i
\end{array}
\right),
\end{equation}
where
$v_i$ is an on-site potential, $\epsilon_F$ the Fermi energy, $\vec h_i$ 
the local exchange field which 
we take in the $xz$ plane, and $\theta_i$ the angle between
$\vec h_i$ and the $z$ axis. The superconducting gap
$\Delta_i=\Delta_0$ in S and $\Delta_i=0$ elsewhere. (This is
appropriate if S is coupled to
the trilayer by a point contact.)

The spin current $\vec J_{i_x}$ is the current associated with the
time derivative of the spin density. The spin current
$J_{i_x,\varepsilon}$ carried by 
a quasi-particle at energy $\varepsilon$ and with
wavefunction $\Psi$ reads
\begin{equation}
\vec J_{i_x,\epsilon} = -t \ {\rm Im } \sum_{i_y,i_z} (\psi^{e\dagger}_{i+d} \vec \sigma  \psi^e_i
- \psi^{hT}_{i+d} \vec \sigma  \psi^{h*}_i),
\end{equation}
with  $d=(1,0,0)$, while the equilibrium
spin current $J_{i_x}^{\rm eq}$ is
\begin{equation}
\vec J_{i_x}^{\rm eq} = -\frac{1}{2} \sum_{\epsilon>0} \vec
J_{i_x,\epsilon}.
  \label{eq:taueq}
\end{equation}
Here,
$\vec \sigma=(\sigma_x,\sigma_y,\sigma_z)^T$ 
is a vector of Pauli matrices acting on the spin
index. Non-conservation of the spin current in F$_a$ and F$_b$ results
in a torque $\vec \tau_a$  ($\vec \tau_b$) exerted on the
magnetic moment of $F_a$ ($F_b$)
(the origin is chosen in the normal metal spacer separating $F_a$ 
and $F_b$),
\begin{equation}
\vec \tau_{a} = \vec J_{-\infty} - \vec J_{0} \ \ ,\ \
\vec \tau_{a} = \vec J_{0} - \vec J_{+\infty}. 
\end{equation}

\section{Non-equilibrium torque}

For a wide trilayer with a diffusive spacer layer, the main
contribution to the non-equilibrium torque lies in the plane
spanned by the two magnetic moments $(\vec h_a,\vec h_b)$
(the $xz$ plane in Fig.\ \ref{fig:schema}).~\cite{wb}
In the absence of the superconducting contact,
the non-equilibrium
torque $\tau_{a(b)}^{\rm ne}$ induced by a current $I$ changes of sign at
$\theta=0,\pi$, thus stabilizing the
configurations $\theta=0$ or $\theta=\pi$ depending of the
direction of the current.~\cite{slon1} The presence of
the superconductor, however, gives a geometric constraint that
allows one to find a relation between $\tau_{a}^{\rm ne}$ and 
$\tau_{b}^{\rm ne}$,
from which it follows that $\tau_a^{\rm ne}$ also vanishes at $\theta=\pi/2$.
To find this relation, note
that the spin current vanishes inside S (for bias voltage
$eV < \Delta_0$). Although the
ferromagnetic layers do not conserve spin current, they do conserve 
the majority and minority currents parallel to their exchange
fields $\vec h_{a(b)}$, hence
\be
\vec \tau_a \perp \vec h_a \ \ , \ \ \vec \tau_b \perp \vec h_b.
\label{eq:perp}
\ee
Since $\vec J_{+\infty}=0$ and $\vec h_b$ points along the $z$-axis, 
we find from Eq.\ (\ref{eq:perp}),
\be
\vec J_{0} = J_0 \ \ \vec x
\ee
Finally, we use that, for wide trilayers, $\vec J_{-\infty}$ is
polarized along $\vec h_a$ and therefore, by Eq.\ (\ref{eq:perp})
gives no contribution to the torque.~\footnote{This is the case 
if the majority and minority
spins are incoherently transmitted or reflected
by the ferromagnetic layers, which
is a good assumption for a wide trilayer with many propagating
channels at the Fermi level. 
However, the conclusion (\ref{eq:tauab}) also holds for coherent 
transmission or reflection Ref.~\ref{wb}.}
The torque $\vec \tau_a$ is thus given by the projection of 
$\vec J_{0}$ to the unit vector
perpendicular to $\vec h_a$ and hence,
\be
\tau_a^{\rm ne}  = \tau_b^{\rm ne} \cos \theta. \label{eq:tauab}
\ee   
Equation (\ref{eq:tauab}) shows that the non-equilibrium torque
$\tau_a^{\rm ne}$ can stabilize the $\theta=\pi/2$
configuration. (Note, however, that the presence of the S contact 
creates an asymmetry between the torques $\tau_a$ and $\tau_b$,
so that it is necessary for this effect
that the magnetic moment of F$_b$ is
held fixed.)

\section{Equilibrium torque}
 
The previous section dealt with the non-equilibrium torque, which
lies in the plane spanned by the two magnetic moments $\vec h_a$ and
$\vec h_b$. It was found that $\tau_a^{\rm ne} = 0$ for
$\theta=\pi/2$. We now ask whether the equilibrium torque has the
same property.

The answer is negative. Again, we can see this by a geometrical
argument. In equilibrium, no spin current flows outside the trilayer
on either side, and thus $\vec \tau_b^{\rm eq} = -\vec\tau_a^{\rm eq}
=\vec J_0$.
In combination with Eq.\ (\ref{eq:perp}), one then finds that
$\vec \tau_b^{\rm eq}$ and $\vec\tau_a^{\rm eq}$ point out of the
plane spanned by $\vec h_a$ and
$\vec h_b$. There is no relation as simple as Eq.\ (\ref{eq:tauab})
for the out-of-plane component of the torque, hence no special
behavior at $\theta=\pi/2$ is expected.

There are two ways to compute the equilibrium torque: either by
a direct calculation of the equilibrium spin currents in the 
trilayer, cf.\ Eq.\ (\ref{eq:taueq}), or from the derivative of the
ground state energy with respect to the angle $\theta$,
\be
\tau_{a}^{\rm eq}=\frac{\partial E}{\partial \theta}.
\ee
It should be emphasized that the equilibrium torque involves 
contributions from energies in the entire
conduction band, as opposed to the non-equilibrium torque where only
quasiparticle states near the Fermi level play a role. 

\begin{figure}
\begin{center}
\psfig{figure=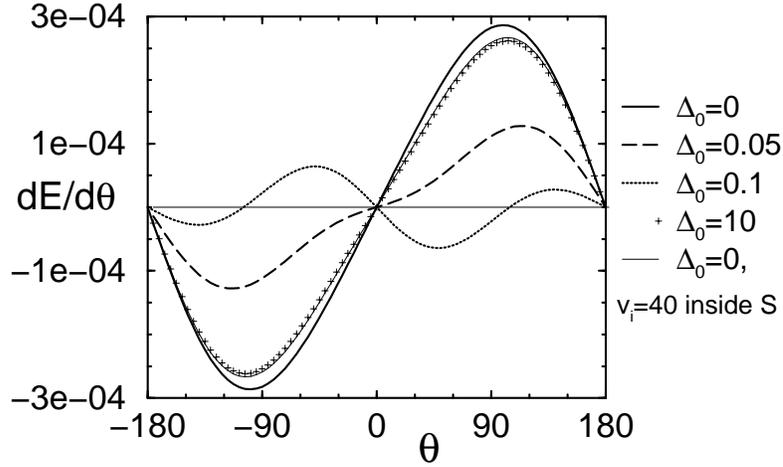,height=3.4in}
\end{center}
\caption{Equilibrium torque $\tau^{\rm eq}$
for a one-dimensional FNF trilayer with
one superconducting contact. Torque is shown as a function of the
angle $\theta$ between the moments of the two ferromagnetic layers,
and for various values the superconducting gap $\Delta_0$, or
without superconducting contact, see figure. The system consists
of $80$ sites, the relative sizes of the different layers being
respectively for N, F$_a$, N, F$_b$, N and S: $9$, $11$, $9$,
$11$, $4$ and $36$. We set $v_i=\epsilon_F=0$ and $h_i=0.8$.  
All energies are measured in units of $t$.
\label{fig:torque}}
\end{figure}

To illustrate the $\theta$ dependence of $\tau_a^{\rm eq}$, we 
have performed a numerical exact diagonalization of the BdG
equation (\ref{eq:BdG}). We have considered the special case of a 
one-dimensional trilayer, $i = i_x$ in Eq.\ (\ref{eq:BdG}).
In Fig.~\ref{fig:torque}, $\tau_{a}^{\rm eq}$ is plotted
as a function of $\theta$. We notice that the equilibrium torque
(exchange interaction) depends strongly on the value of the 
superconducting gap $\Delta_0$. This implies that both 
the direct spin exchange between the two
magnets and other contributions that involve the superconductor
play a role. For a very large gap $\Delta_0 \gg t$, where $2t$
is the width of the conduction band in N, the superconductor acts as a
hard wall and the results are qualitatively the same as without
superconductivity. However, an interesting regime appear for
relatively small values of the gap $\Delta_0 < t$ where the 
equilibrium torque depends strongly on $\Delta_0$, although the 
actual $\theta$-dependence of $\tau_a^{\rm eq}$ is very sensitive 
to the particular choice of parameters and, for a realistic 
calculation, would require knowledge of the
detailed band structure of the different materials. 
In the particular example shown in Fig.~\ref{fig:torque},
for $\Delta_0=0.1 t$ the biquadratic harmonic 
dominates the exchange coupling~\cite{slon3}, showing that, after
eventual fine tuning of parameters, nontrivial values of $\theta$
can be stabilized. (A similar effect may also occur without a
superconducting contact, see, e.g., Ref.\ \ref{slon3}.)

\section*{Acknowledgments}
We thank P.\ Chalsani, A.\ A.\ Clerk, 
E.\ B.\ Myers, and D.\ C.\ Ralph for useful discussions.
This work was supported by the
NSF under grant no.\ DMR 0086509 and by the Sloan foundation.

\end{document}